\documentclass[aps,prx,twocolumn,superscriptaddress,amsmath,amssymb,nofootinbib]{revtex4-2}
\usepackage{graphicx,enumitem}
\usepackage{standalone}
\usepackage{multirow,dcolumn}
\usepackage{txfonts}
\usepackage[hyperindex,breaklinks]{hyperref}
\usepackage{soul,cancel}
\usepackage{epsf,amsmath,amssymb,verbatim,color,multirow,pifont,graphicx,cleveref,tabularx}
\usepackage[dvipsnames]{xcolor}
\usepackage{rotating}
\usepackage{amsmath,amsfonts,amssymb,bm,cancel}

\begin{document}
\title{
Effective vaccination strategy using graph neural network ansatz}
	
\author{Bukyoung Jhun}
\email{jhunbk@snu.ac.kr}
\affiliation{CCSS, CTP and Department of Physics and Astronomy, Seoul National University, Seoul 08826, Korea}

\begin{abstract}
The effectiveness of vaccination highly depends on the choice of individuals to vaccinate, even if the same number of individuals are vaccinated.
Vaccinating individuals with high centrality measures such as betweenness centrality (BC) and eigenvector centrality (EC) are effective in containing epidemics.
However, in many real-world cases, each individual has distinct epidemic characteristics such as contagion, recovery, fatality rate, efficacy, and probability of severe reaction to a vaccine.
Moreover, the relative effectiveness of vaccination strategies depends on the number of available vaccine shots.
Centrality-based strategies cannot take the variability of epidemic characteristics or the availability of vaccines into account.
Here, we propose a framework for vaccination strategy based on graph neural network ansatz (GNNA) and microscopic Markov chain approach (MMCA).
In this framework, we can formulate an effective vaccination strategy that considers the properties of each node, and tailor the vaccination strategy according to the availability of vaccines.
Our approach is highly scalable to large networks.
We validate the method in many real-world networks for network dismantling, the susceptible-infected-susceptible (SIS) model with homogeneous and heterogeneous contagion/recovery rates, and the susceptible-infected-recovered-dead (SIRD) model.
We also extend our method to edge immunization strategy, which represents non-pharmaceutical containment measures such as travel regulations and social distancing.
\end{abstract}

\maketitle

\section{Introduction}

Epidemics do not occur randomly; instead, they spread through structured interactions among the host population.
Network theory provides an integrated framework to study the effects of the structure of interactions on dynamical processes~\cite{Dorogovtsev2002,Barabasi2009,Boccaletti2006,Boccaletti2014,Battiston2020}.
For epidemic processes, individuals are represented as nodes, and contacts between individuals are represented as edges (links) in the network.
Traditional theories of epidemic spreading ignored network effects~\cite{Anderson1992,Keeling2011}; however, extensive research devoted to network epidemiology demonstrated that the structural properties of network such as heterogeneity of degree (number of edges a node has) significantly affect the spreading of epidemics~\cite{Pastor-Satorras2015,Pastor-Satorras2001,Pastor-Satorras2001a,Moreno2002,Ferreira2012}.
Such network effects have significant implications because most real-world social systems exhibit highly complex connectivity patterns characterized by heavy-tailed distributions~\cite{Dorogovtsev2002,Barabasi2009,Boccaletti2006,Boccaletti2014}.
Network epidemiology has also been applied to social spreading processes such as the spread of innovations, rumors, and opinions~\cite{Moreno2004a,Watts2007,Katona2011,Fernandez-Gracia2014}.

Containing, mitigating, and preventing the spread of epidemics is a crucial goal in mathematical epidemiology, therefore,  extensive research has been devoted to developing effective vaccination strategies in complex networks~\cite{Pastor-Satorras2015,Wang2016,Wang2017,Pastor-Satorras2002,Cohen2003,Madar2004,Chen2008,Schneider2011,VanMieghem2011,Hebert-Dufresne2013,Clusella2016,Matamalas2018,Jhun2021}.
Effective vaccination strategies aim to vaccinate the optimal set of nodes in the network to minimize the damage caused by epidemic diseases such as the total number of infections or epidemic mortality.
It has been found that the effectiveness of a vaccination highly depends on which nodes we choose to vaccinate even if we choose the same number of nodes.
This problem is relevant to the current situation where the number of effective SARS-CoV-2 vaccine shots is less than the total population in most countries, especially in developing countries~\cite{Tatar2021}.
Moreover, the vanishing epidemic threshold of scale-free networks~\cite{Pastor-Satorras2001,Pastor-Satorras2001a} suggests that such pandemic will presumably occur repeatedly; therefore, it is crucial to be prepared for another vaccine shortage.

Graph neural networks (GNNs) are deep learning--based methods that operate on graphs or networks where other types of machine-learning methods such as convolutional neural networks (CNNs) or recurrent neural networks (RNNs) cannot be implemented because of the irregular and non-Euclidean nature of the complex network.
GNN has become a widely used method for network analysis because of its convincing performance in various fields, such as estimation of molecular properties~\cite{Yuan2021,Hao2020}, drug discovery~\cite{Gomez-Bombarelli2018}, and traffic forecasting~\cite{Bui2021,Roy2021}.
In the epidemic field, GNNs have been employed for the prediction of disease prevalence~\cite{Murphy2020,Panagopoulos2020,Deng2020}, identification of patient zero~\cite{Shah2020}, and estimation of epidemic state using limited information~\cite{Tomy2021}.
Few studies have developed dynamic epidemic control schemes that identify epidemic hotspots from the partially observed epidemic state of each individual~\cite{Song2020a,Meirom2020}.

Here, we propose a framework for vaccination strategy in complex networks based on GNN.
By employing graph neural network ansatz (GNNA) and microscopic Markov chain approach (MMCA), we can determine the optimal strategy through few mean-field calculations.
Note that comparing the performances of two similar vaccination strategies generally requires an excessive number of Monte Carlo epidemic simulations. 
This framework can be implemented to formulate effective vaccination strategies, tailored to the available amount of vaccine shots, for various epidemic processes in a complex network.
If the properties of each node, such as contagion, recovery, or fatality rate, are distinct, the GNNA can systematically consider this information to formulate an optimal strategy.
Such a situation wherein the nodes of a network possess distinctive characteristics is relevant in real-world epidemics.
For instance, the case fatality rate of COVID-19 varies significantly according to age~\cite{Levin2020,Manuel2020,Kim2020,Barone-Adesi2020,Li2020,Shim2021a,Bhatt2021}; hence, it is not trivial to determine whether senior population with high fatality rate or young population with high contact rate should be primarily vaccinated to minimize the epidemic mortality~\cite{Jhun2021a}.
The age-dependent efficacy and probability of severe reaction to vaccines further complicate the issue~\cite{Reingold1985,Kumar2008}.
However, there has been no vaccination strategy that systematically takes the epidemic properties of each node into account.

To prove the validity of our algorithm, we test this method for network dismantling, the susceptible-infected-susceptible (SIS) model with homogeneous and heterogeneous epidemic parameters, and the susceptible-infected-recovered-dead (SIRD) model in many real-world networks with up to 320K nodes and 1M edges.
We also extend this framework to edge immunization, which represents non-pharmaceutical epidemic containment measures such as travel regulations and social distancing.
We compare the performance of the proposed framework with the existing centrality-based methods.
The proposed method outperforms the centrality-based vaccination strategies at all levels of vaccine supply.
Moreover, because GNNA considers the properties of each node and tailors the vaccination strategy to the specific amount of vaccine available, it allows us to find new phenomena such as the transition of optimal strategies from high-fatality to high-BC strategies according to the level of vaccine supply.

\section{Model}

\subsection{SIS model with homogeneous and heterogeneous contagion/recovery rate}

The SIS model is one of the most extensively studied epidemic models in complex network~\cite{Pastor-Satorras2015,Pastor-Satorras2001a,Ferreira2011,Ferreira2012,Matamalas2018,Iacopini2019,Jhun2019,Battiston2021}.
Recently, the SIS model where the recovery rate varies from node to node was introduced~\cite{DeArruda2020}.
We extend this model and let the contagion rate that each node infects others to be varying as well.
Such variability of epidemic parameters is a natural assumption because the prognosis of an epidemic disease depends on the age and other factors of each individual.

In the SIS model, each node is either in the susceptible (S) or infected (I) state.
At each time step, each infected node $j$ infects its neighbors with probability $\beta_j$, hence susceptible node $i$ turns into the infected state with probability
\begin{align}
	P_i = 1-\left(1-\beta_{j}\right)^{n_i^\mathrm{I}} \,,
\end{align}
where $n_i^\mathrm{I}$ is the number of infected neighbors of node $i$.
An infected node turns into the susceptible state with probability $\mu_i$.
If $\beta_i=\beta$ and $\mu_i=\mu$, the equation is reduced to the traditionally studied SIS model with homogeneous contagion and recovery rate.
For heterogeneous cases, the distribution of the contagion and recovery rates can be arbitrary, but in this study, the contagion rate and recovery rate of node $i$ are uniformly distributed between $0$ and $\beta$, and $0$ and $\mu$, respectively.

We start the simulation with the fully infected system and evolve the system for $t_\mathrm{relax}=2\times10^4$ so that the system reaches its stationary state.
Then, the density of infection is sampled for $t_\mathrm{sample}=2\times10^4$.
Quasistationary method~\cite{Ferreira2011,Ferreira2012} and other approaches~\cite{Jo2020} have been employed to obtain the steady-state of the epidemic dynamics in previous studies.
Here, we apply a small conjugated field $h_i=10^{-3}\mu_i$ on each node $i$ to keep the system in the active state~\cite{Lubeck2004}.
The intensity of the conjugated field is irrelevant as long as the value is very low.

\subsection{SIRD model}

Although mortality is one of the significant damage caused by epidemic diseases, the susceptible-infected-recovered (SIR) model cannot be used to study the vaccination strategy to minimize epidemic mortality, because recovery and death are not distinguished in the SIR model.
The SIRD model was therefore introduced as a minimal epidemic model to study epidemic mortality.

In the SIRD model, each node is in either susceptible (S), infected (I), recovered (R), or dead (D) states.
The infection occurs by the same rule with the SIS model.
A susceptible node turns into the infected state with probability
\begin{align}
	P_i = 1-\left(1-\beta\right)^{n_i^\mathrm{I}} \,,
\end{align}
where $n_i^\mathrm{I}$ is the number of infected neighbors of node $i$.
At each time step, an infected node turns to the R state with probability $(1-\mathrm{IFR}_i)\mu$ and to D state with probability $\mathrm{IFR}_i\cdot\mu$.

Recovery and death occur with ratio $(1-\mathrm{IFR}_i):\mathrm{IFR}_i$, therefore, the infection fatality rate (IFR) of node $i$ is $\mathrm{IFR}_i$.
IFR is defined as the ratio of deaths caused by disease to the total number of people infected with the disease.
The fatality rate of epidemic diseases such as COVID-19 significantly depends on age and other morbidity factors~\cite{Li2020a,Jordan2020}.
Therefore, it is important to study the SIRD model where the fatality rate varies from node to node.
We start the simulation after infecting a small fraction $10^{-3}$ of nodes in the network.
All the reactions (infection, recovery, and death) include an infected node; therefore, if the number of infected nodes becomes zero, then the epidemic dynamics ends.
In this study, we sampled the mortality rate for $n_\mathrm{sample}=2\times10^4$.

\subsection{Construction of a multiplex network from contact data}

To investigate the effectiveness of the vaccination strategies on real-world epidemic diseases, we construct a multiplex network from human contact patterns between age groups and the degree distribution, and the age-dependent IFR of COVID-19 was implemented.
The network is constructed from the contact matrix $M_{\alpha\beta}$, which is the average number of contacts that an individual in group $\alpha$ has with individuals in group $\beta$, obtained by survey~\cite{Mistry2021}.
The human contact degree distribution follows negative binomial distribution $\mathrm{NB}\left(r, p\right)$ with $r \simeq 0.36$~\cite{Mossong2008}.
The parameter $p_\beta$ of age group $\beta$ is determined by the average degree $\left<k\right>_\beta = \sum_\alpha M_{\alpha\beta}$: $p_\beta = 1-\left<k\right>_\beta / \left(r+\left<k\right>_\beta\right)$.

The data was collected for people of age 0 to 84, and people of age 85 and above were aggregated.
We extend the data to people of age 99 by assuming that people of age 85 and above exhibit identical contact patterns.
First, we draw the degree of each node from the degree distribution $\mathrm{NB}\left(r, p_\beta\right)$ of the corresponding age group, and place "stubs" of that number.
We then select a stub with equal probability and connect it with another stub, which is selected with probability proportional to $M_{\alpha\beta}$, where $\alpha$ is the age group of the first selected stub and $\beta$ is that of the second selected stub.
This iteration is repeated until only one or no stub is left (If only one stub is left, it cannot be matched with any other stub).

The IFR of each node is calculated based on meta-analysis of medical literature~\cite{Levin2020}, where the age dependent IFR is calculated as
\begin{align}
    \log_{10} {\rm IFR} = (-3.27 \pm 0.07) + (0.0524 \pm 0.0013) \, {\rm age} \,.
\end{align}

\section{Vaccination strategy}

\subsection{Graph neural network ansatz (GNNA)}

We aim to vaccinate the optimal set of $q$ nodes to minimize the damage caused by an epidemic process, such as the total number of infections or infectious deaths.
A vaccinated node does not get infected even if it has contact with infected nodes.
We suppose that each node's fitness to be vaccinated in the network can be expressed by an $L$-layered GNN, namely GNNA.
\begin{align}
    \boldsymbol{s}_{i}^{(\ell)}	=\mathrm{Agg}\left(\boldsymbol{s}_{i}^{(\ell-1)},\cup_{j\in\mathrm{n.n.\,of\,}i}\boldsymbol{s}_{j}^{(\ell-1)}\right) \,,
\end{align}
and
\begin{align}
    \boldsymbol{s}_{i}^{(0)} &= \boldsymbol{x}_{i} \,,\\
    \boldsymbol{z}_{i} &= \boldsymbol{s}_{i}^{(L)} \,,
\end{align}
where $\mathrm{Agg}$ is the function that aggregates the information from the neighbors of each node, $\boldsymbol{x}_i$ is the vector of node features of node $i$ such as its contagion rate, recovery rate, fatality rate, efficacy of vaccine, or probability of having a severe reaction to the vaccine.
$\boldsymbol{s}_i^{(\ell)}$ is the vector of hidden state of node $i$ in layer $\ell$, and $\boldsymbol{z}_{i}$ is the output of the GNN.	
Various functions have been used for the aggregation function~\cite{Kipf2017,Hamilton2017,Velickovic2017}.
Here, we take the form
\begin{align}
    s_{i}^{(0)}&=\sigma\left(w_{0}^{(0)}+w_{1}^{(0)}x_{i1}+w_{2}^{(0)}x_{i2}+\cdots\right) \label{eq:gnna_1} \,,\\
    s_{i}^{(1)}&=\sigma\left(w_{0}^{(1)}+w_{1}^{(1)}s_{i}^{(0)}+w_{2}^{(1)}k_{i}^{w_{4}^{(1)}-1}\sum_{j\in\mathrm{n.n.\,of\,}i}(s_{j}^{(0)}+w_{3}^{(1)})\right) \,,\\
    &\vdots \nonumber \\
    s_{i}^{(L)}&=\sigma\left(w_{0}^{(L)}+w_{1}^{(L)}s_{i}^{(L-1)}+w_{2}^{(L)}k_{i}^{w_{4}^{(2)}-1}\sum_{j\in\mathrm{n.n.\,of\,}i}(s_{j}^{(L-1)}+w_{3}^{(L)})\right) \label{eq:gnna_3} \,,
\end{align}
where $k_i$ is the degree of node $i$, and we choose leaky rectified linear unit (ReLU) for the activation function $\sigma(\cdot)$.
The introduction of $w_4^{(\ell)}$ allows GNNA to include both the summation ($w_4^{(\ell)}=1$) and average ($w_4^{(\ell)}=0$) for the aggregation.
The permutational invariance among neighboring nodes is retained.
In this study, we use $L=2$.
The output $s_i^{(L)}=s_i^{(2)}$ is the fitness of node $i$ to be vaccinated.	
This fitness effectively works as a centrality measure tailored to the epidemic process of the interest.
We vaccinate $q$ nodes with the highest fitness.

The ansatz Eqs.~\eqref{eq:gnna_1}--\eqref{eq:gnna_3} includes various vaccination strategies.
For instance, if 
\begin{align}
    w_{r_0}^{(0)} &= 1 \,, \\
    w_{r}^{(0)} &= 0 \quad \forall r\neq r_0 \,, \\
    w_{1}^{(\ell)} &= 1 \quad \forall \ell>0 \,, \\
    w_{r}^{(\ell)} &= 0 \quad \forall r\neq 1, \ell>0 \,,
\end{align}
the fitness of each node becomes equal to its node feature $x_{ir_0}$, and we vaccinate the nodes in descending order of their feature (fatality rate of the node, for instance).
If 
\begin{align}
    w_{0}^{(0)} &= 1 \,, \\
    w_{r}^{(0)} &= 0 \quad \forall r>0 \,, \\
    w_{0}^{(1)} &= w_{1}^{(1)}=w_{3}^{(1)}=0 \,, \\
    w_{2}^{(1)} &= w_{4}^{(1)}=1 \,, \\
    w_{1}^{(\ell)} &= 1 \quad \ell>1 \,, \\
    w_{r}^{(\ell)} &= 0 \quad \forall r\neq 1, \ell>1 \,,
\end{align}
the fitness equals the degree of each node.
Other strategies such as averaging the node features of the nearest or second-nearest neighbors of a node can be represented by the Eqs.~\eqref{eq:gnna_1}--\eqref{eq:gnna_3}.

Because the weights $w_r^{(\ell)}$ are shared over the entire network, the number of parameters of GNNA is $5L+m+1$, where $m$ is the number of node features.
Moreover, the actual dimension of the manifold represented by GNNA is lower.
The output of GNNA, which is the vaccination strategy, is invariant under the following transforms for each $0 \leq \ell \leq L$ (because if the input of leaky ReLU scales by a factor of $\alpha$, so does the output):
\begin{align}
w_r^{(\ell)} &\rightarrow \alpha w_r^{(\ell)} \quad \forall r \,,\\
w_0^{(\ell^\prime)} &\rightarrow \alpha w_0^{(\ell^\prime)} \quad \forall \ell^\prime > \ell \,,\\
w_3^{(\ell^\prime)} &\rightarrow \alpha w_3^{(\ell^\prime)} \quad \forall \ell^\prime > \ell \,,
\end{align}
while other weights are kept unchanged.
Additionally, because we are only interested in the rank of $s_i^{(L)}$, the parameter $w_0^{(L)}$ is irrelevant.
The dimension of the manifold is, therefore, $4L+m$.
Moreover, when there is no node feature, $w_3^{(1)}$ becomes irrelevant because $s_j^{(0)}$ is constant, and the dimension is $4L-1$.
Therefore, GNNA is highly scalable to large networks.

We can extend GNNA to edge immunization by aggregating the fitness of nodes in each edge. 
Because an edge is always connected to two nodes, we employ two-dimensional Taylor series expansion for the aggregation function.
The fitness $s_{(i,j)}$ of an edge $(i,j)$ is then
\begin{align}
    s_{(i,j)} &= w_0^{(L+1)} \left(s_i^{(L)} + s_j^{(L)}\right) + w_1^{(L+1)} \left(s_i^{(L)2} + s_j^{(L)2}\right) \nonumber\\
    &+ w_2^{(L+1)} s_i^{(L)} s_j^{(L)} + \cdots \,,
\end{align}
where the coefficients are chosen so that there is a symmetry between $i$ and $j$.
Here, we only use quadratic terms $w_0^{(L+1)}$, $w_1^{(L+1)}$, and $w_2^{(L+1)}$.
Therefore, three additional parameters are required for the edge immunization.

\subsection{Microscopic Markov chain approach (MMCA)}

The stochasticity of the epidemic processes brings a challenge to the optimization problem.
Because of the fluctuation in the results of the epidemic simulations, the average of the sampled density of infection can exhibit a low value even if the expectation value is not low.
If the gradient descent method is directly implemented, the trajectory of the optimization may be forever affected by a low value once obtained because of the fluctuation.
Also, to compare the performances of two similar vaccination strategies, which likely have similar expectation values of the density of infection, an excessive number of Monte Carlo simulations have to be performed.
A point with high fluctuation can be selected as the optimal point even if the expectation value is not low.

To avoid such issues, we employ MMCA~\cite{Gomez2010,Gomez2011,Matamalas2020}, to analytically estimate the performance of the vaccination strategies.
MMCA solves the mean-field equation for each node in the network to provide more accurate predictions of the epidemic prevalence than heterogeneous mean-field (HMF) theory~\cite{Pastor-Satorras2001,Pastor-Satorras2001a,Moreno2002}.
Because there is no fluctuation in the result of the MMCA, the aforementioned problem can be avoided.
We show that even when the GNNA is optimized with MMCA, the resulting vaccination strategies effectively minimize the density of infection of the stochastic epidemic model.

The MMCA tracks the probability $P_{i}^\mathrm{X}(t)$ of each node $i$ being in state $\mathrm{X}$ at time $t$~\cite{Gomez2010,Gomez2011,Matamalas2020}.
For SIS model, we track $P_{i}^\mathrm{I}(t)$.
The MMCA equations of the SIS model with heterogeneous contagion/recovery rate is expressed
\begin{align}
    P_{i}^\mathrm{I}(t+1) = P_{i}^\mathrm{S}(t)\left[1-\prod_{j\in \mathrm{n.n.\,of\,}i}\left(1-\beta_{j}P_{j}^\mathrm{I}(t)\right)\right]+\left(1-\mu_{i}\right)P_{i}^\mathrm{I}(t) \label{eq:sis} \,,
\end{align}
and $P_i^\mathrm{S}(t)=1-P_i^\mathrm{I}(t)$.
For the traditionally studied SIS model with homogeneous contagion and recovery rate, $\beta_i=\beta$ and $\mu_i=\mu$.
We solve Eq.~\eqref{eq:sis} for its fixed point to determine the stationary state.

For SIRD model, we track $P_{i}^\mathrm{I}(t)$, $P_{i}^\mathrm{R}(t)$, and $P_{i}^\mathrm{D}(t)$.
The MMCA equations of the SIRD model is expressed
\begin{align}
    P_{i}^\mathrm{I}(t+1) &= P_{i}^\mathrm{S}(t)\left[1-\prod_{j\in \mathrm{n.n.\,of\,}i}\left(1-\beta P_{j}^\mathrm{I}(t)\right)\right]+\left(1-\mu\right)P_{i}^\mathrm{I}(t) \label{eq:sird_1} \,,\\
    P_{i}^\mathrm{R}(t+1) &= (1-\mathrm{IFR}_i) \mu P_{i}^\mathrm{I}(t) \,,\\
    P_{i}^\mathrm{D}(t+1) &= \mathrm{IFR}_i \cdot \mu P_{i}^\mathrm{I}(t) \,, \label{eq:sird_3}
\end{align}
and $P_i^\mathrm{S}(t)=1-P_i^\mathrm{I}(t)-P_i^\mathrm{R}(t)-P_i^\mathrm{D}(t)$.
We solve Eqs.~\eqref{eq:sird_1}--\eqref{eq:sird_3} until $\sum_i P_{i}^\mathrm{I}(t) < \epsilon = 10^{-4}$, then calculate the mortality rate $\sum_i P_{i}^\mathrm{D} / N$.

\subsection{Gaussian random walk--based optimization}

\begin{figure}
    \centering
	\includegraphics[width=\columnwidth]{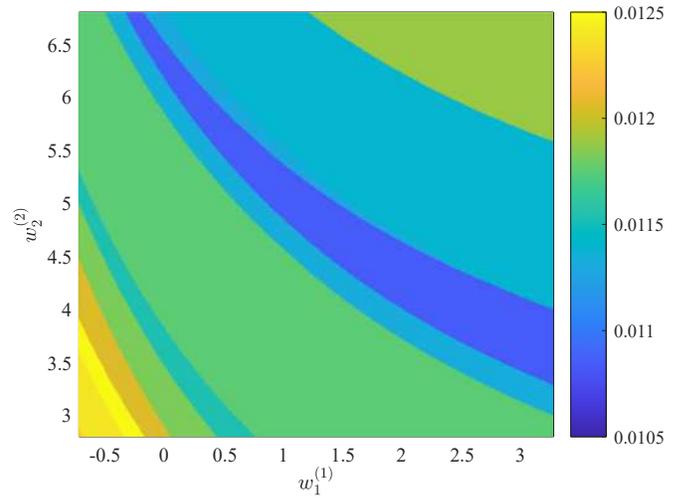}
	\caption{
	Density of infection of SIS model calculated by MMCA as the function of weight parameter $w_r^{(\ell)}$ of GNNA.
	The plot illustrates the loss surface projected on the $w_1^{(1)}$-$w_2^{(2)}$ plane.
	The value is flat everywhere except on certain lines.
	The flat regions each correspond to an identical vaccination strategy; therefore, the objective function does not vary in the region.
	The vaccination is tested in the airline network with the contagion rate $\beta=0.2$, recovery rate $\mu=0.5$, and vaccination rate $q/N=0.1$.
    }
	\label{fig:landscape_loss}
\end{figure}

\begin{figure*}
	\includegraphics[width=\textwidth]{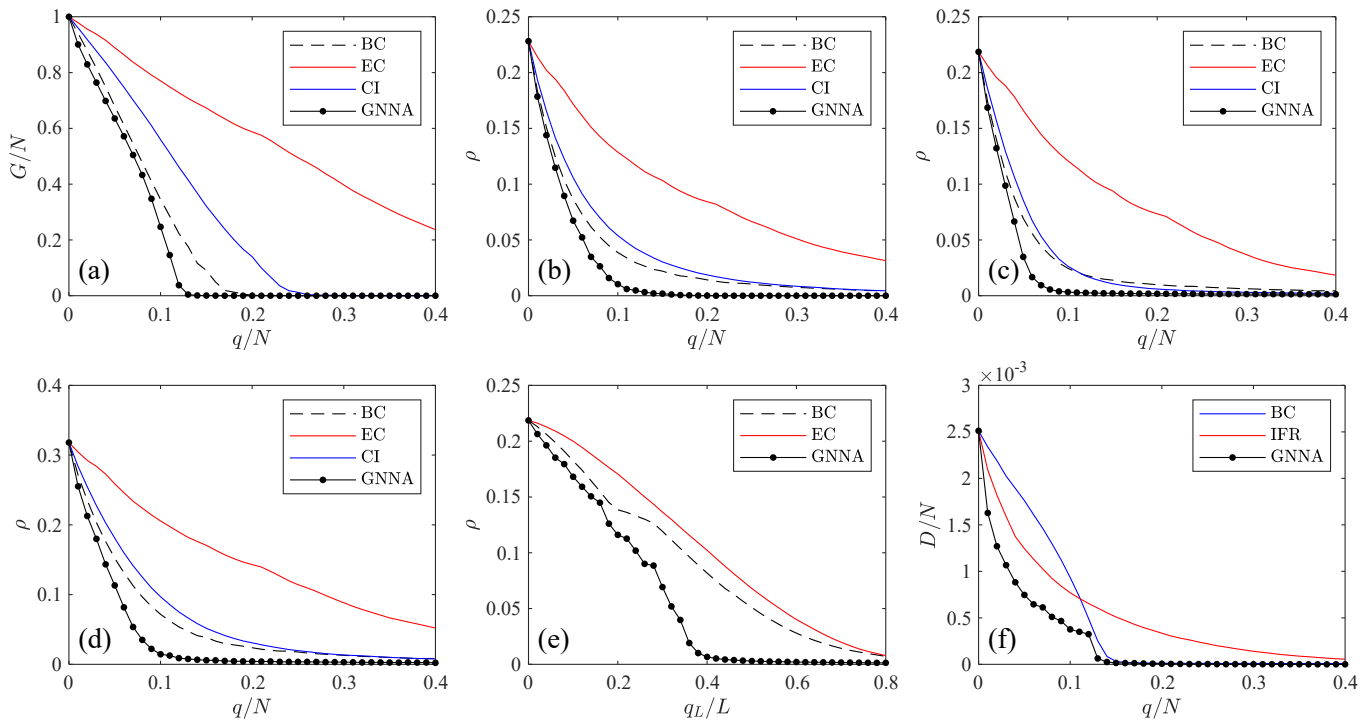}
	\caption{
	Performance of GNNA compared to centrality-based vaccination strategies.
	(a) Largest connected component size of network dismantling and the density of infection $\rho$ of (b) MMCA SIS model, (c) SIS model, (d) SIS model with heterogeneous contagion and recovery rate, (e) edge immunization of SIS model, and (f) mortality rate $D/N$ of SIRD model.
	The number of nodes in the network is $N$, the number of edges is $L$, the number of vaccinated nodes is $q$, and the number of immunized edges is $q_L$.
	The vaccination strategy obtained by GNNA outperforms all centrality measures at all vaccination levels.
    }
	\label{fig:performance}
\end{figure*}

The loss surface of GNNA is distinct from the loss landscape of usual neural networks~\cite{Li2018}.
The loss, which is the density of infection calculated by MMCA in this case, is illustrated in Fig.~\ref{fig:landscape_loss} for the SIS model.
It is flat almost everywhere; at certain lines, there is a leap.
This is because the parameters of the GNNA are continuous but the vaccination strategy is discrete.
For a small perturbation of the weights, except for special cases, the vaccination strategy formulated by the GNNA is invariant, as is the objective function.
Therefore, the gradient descent and other gradient descent--based optimization algorithms such as SQP~\cite{Gill2005} and the Nelder--Mead method~\cite{Nelder1965} cannot be used in this case.

GNNA reduces the exponentially large dimension of the space of the vaccination to 7--10 and allows Gaussian random walk to effectively optimize the vaccination strategies.
Initially, the weights of the provisional solution are set as $w_0^{(0)}=1$, $w_1^{(\ell)}=1$ for all $\ell>0$, and zero if otherwise.
This way, the fitnesses of all nodes are equal to one.
At each step, we perturb the weights $w_r^{(\ell)} \rightarrow w_r^{(\ell)} + \omega_r^{(\ell)}$, where $\omega_r^{(\ell)}$ independently follows the Gaussian distribution with zero mean: $\omega_r^{(\ell)} \sim \mathcal{N}(0,\sigma^2)$.
The standard deviation $\sigma$ is initially $0.5$ and decreases by a factor of $1 - 5/n_\mathrm{iter}$ at each step (the standard deviation becomes 0.003 at the end of the iteration).
This way, as the iteration progresses, we can focus on finding a more detailed position of the minimum in the loss landscape.
There is a probability that the perturbed weight returns the same set of nodes as the provisional solution.
In such cases, we find another position without calculating the objective function again (Because the result is the same as the provisional solution).

\section{Results}

\subsection{Effectiveness of the vaccination strategy}

Vaccinating nodes with high centrality measures has been reported to effectively reduce epidemics in complex networks.
One example of such centrality measure is \textit{betweenness centrality} (BC), or \textit{load}~\cite{Newman2001,Goh2001a}.
This measure is related to the number of shortest paths passing through the node or edge.
It was found that immunizing nodes or edges with high BC is effective in containing epidemics~\cite{Schneider2011,Cohen2003}.
However, the computational complexity for calculating the BC is $O(N^2 \log N)$.
This significantly limits its capability to be used in large networks.

As an alternative to BC, collective influence (CI) was introduced~\cite{Morone2015}.
The CI provides a scalable centrality measure that considers the local stability of message-passing equations.
Vaccinating (or eliminating) nodes with the highest CI leads to effective dismantling (or herd immunity) of a network.
The collective influence of a node $i$ is
\begin{align}
    C_{\ell}(i)=\left(k_{i}-1\right)\sum_{j\in\partial\mathrm{Ball}(i,\ell)}\left(k_{j}-1\right) \,,
\end{align}
where $\partial\mathrm{Ball}(i,\ell)$ is the set of nodes that have distance $\ell$ from node $i$ (surface of a ball with radius $\ell$).
Although the algorithm becomes exact as $\ell\rightarrow \infty$ for treelike networks, a small $\ell$ yields good results in general complex networks.
In this study, we take $\ell=2$.
CI can be calculated within a time complexity of $O(N\log N)$.
Eliminating nodes with high CI effectively reduces the size of the largest connected component in the network and contains epidemics.

Further, vaccinating nodes with high eigenvector centrality (EC), which has a computational complexity of $O(N\log N)$, is effective in reducing epidemics~\cite{VanMieghem2011}.
Other centrality measures such as K-core index~\cite{Klemm2012}, closeness~\cite{Chen2012}, K-shell~\cite{Kitsak2010}, and H-index~\cite{Drewniak2014} have been used to formulate vaccination strategies; however, one cannot conclude which of these strategies is the most effective because the efficiency of the strategies varies depending on the network and level of vaccine supply.
In this study, we show the performance of BC, which is believed to be effective in a wide class of networks~\cite{Pastor-Satorras2015,Wang2016,Wang2017}, and CI as benchmarks.
Recalculating these centrality measures after each node vaccination enhances the performance of the vaccination; however, this increases the time complexity of the algorithm by a factor of $N$, and makes the method no longer scalable.
Other vaccination strategies that can be implemented when the entire network structure is not available have been researched~\cite{Cohen2003,Salathe2010,Dong2020}; however, these strategies are not as effective as the centrality-based methods.

We tested and compared GNNA-based vaccination with centrality-based strategies for network dismantling, SIS models with homogeneous and heterogeneous contagion/recovery rate, edge immunization of SIS model, and SIRD model.
The strategies were tested in various networks~\cite{Guimera2003,Yang2015,Matamalas2018,Lee2021}, with the number of nodes ranging from 1K to 320K and number of edges from 5K to 1M (the specifics of the networks are provided in the Supplementary Table S1).
We only show the result from one network for each epidemic process (a multiplex network constructed from the human contact pattern for the SIRD model and DBLP Coauthorship network for the others); the rest is provided in Supplementary Figures S2--S6.

Network dismantling is a problem of finding an optimal set of $q$ nodes that breaks the largest connected component of a network into small components with \textit{subextensive} size.
It can be mapped to the optimal vaccination strategy for the spreading process~\cite{Morone2015,Clusella2016}.
For the network dismantling, we directly calculate the size of the largest connected component instead of employing MMCA because there is no stochasticity in this process.
The size of the largest connected component of the DBLP Coauthorship network dismantled by GNNA is illustrated in Fig.~\ref{fig:performance}(a).
For comparison, we plotted the performance of the BC-, EC-, and CI-based strategies.
GNNA-based strategy outperforms all centrality-based strategies at all vaccination levels.

\begin{figure}[!htp]
	\includegraphics[width=\columnwidth]{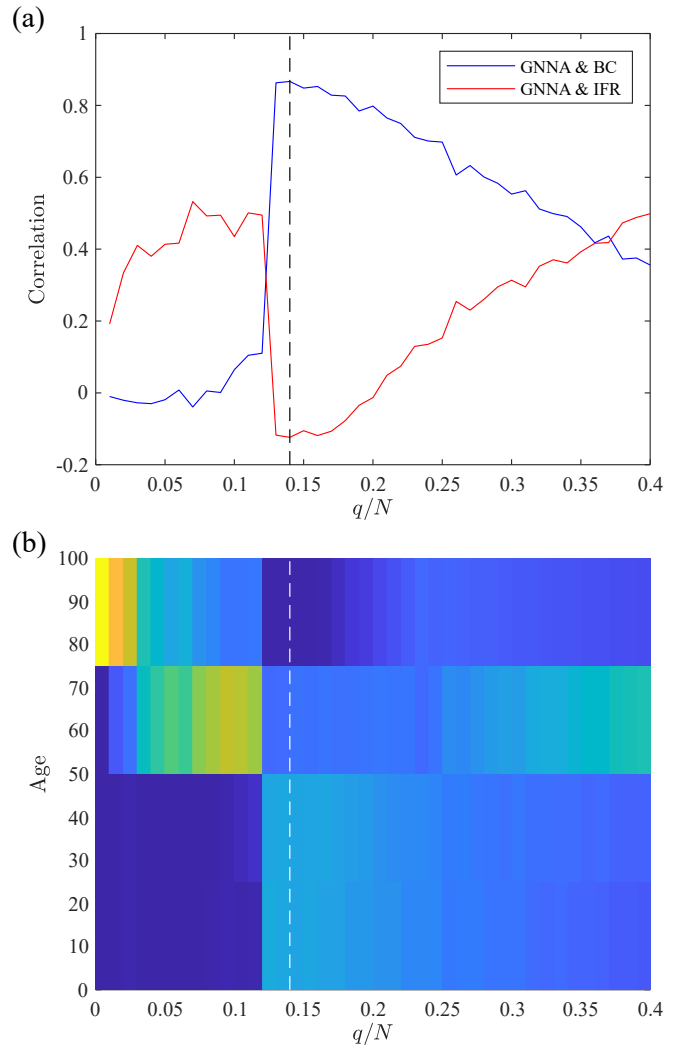}
	\caption{
	Transition of the optimal vaccination strategy in the SIRD model.
	(a) Phi coefficient between the nodes vaccinated by GNNA and high-BC/IFR vaccination strategies.
	The point where herd immunity is achieved by GNNA-based strategy (but not necessarily by other methods) is depicted by the dashed black line.
	When the total vaccination rate $q/N$ is low, the set of nodes vaccinated by GNNA has a large overlap with the high-IFR strategy.
	As the vaccination level approaches the state where herd immunity is possible, GNNA adjusts its strategy, wherein it becomes similar to the high-BC strategy.
	(b) Vaccination rate of four age groups when GNNA-based vaccination is applied.
	Yellow, green, and blue represent high, moderate, and low vaccination rates, respectively.
	When the total vaccination rate $q/N$ is low, the senior age group whose IFRs are the highest is primarily vaccinated.
	However, when the vaccination rate becomes high and approaches herd immunity, which is depicted by the dashed white line, the individuals below age 50 who have high contact rates are primarily vaccinated.
    }
	\label{fig:correlation}
\end{figure}

Further, we tested the performance of GNNA-based vaccination for the SIS model.
The density of infection calculated by MMCA is illustrated in Fig.~\ref{fig:performance}(b), and the result of the Monte Carlo simulation is illustrated in Fig.~\ref{fig:performance}(c).
GNNA-based strategy outperforms all centrality-based strategies at all vaccination levels.
The results for the SIS model with heterogeneous contagion and recovery rates are illustrated in Fig.~\ref{fig:performance}(d).
The disparity between the performance of GNNA-based strategy and centrality-based strategies is greater than the homogeneous case because GNNA considers the epidemic properties of each node whereas centrality-based methods do not.

For edge immunization, vaccinating edges with high edge BC or high edge EC is effective~\cite{Schneider2011,Matamalas2018}.
The edge EC is calculated as the product of the ECs of the two nodes in the edge.
It has been shown that iteratively eliminating edges with the highest link epidemic importance is effective~\cite{Matamalas2018}; however, the complexity of the algorithm is $O(N^2)$. 
The results of the edge immunization are illustrated in Fig.~\ref{fig:performance}(e).
The performances of high edge BC and high edge EC vaccinations are plotted as benchmarks.
GNNA-based strategy outperforms all the edge centrality-based methods at all vaccination levels.

For the SIRD model, the results are illustrated in Fig.~\ref{fig:performance}(f).
Vaccinating nodes with high IFR is effective in reducing the number of deaths; hence, a high-IFR vaccination strategy has been employed in many countries to minimize the mortality due to COVID-19.
The performances of high-BC and high-IFR strategies are shown as baselines.
The high-IFR strategy is more effective than the high-BC strategy when the vaccination rate is low; however, the high-BC strategy outperforms the high-IFR strategy when the vaccination rate is high.
GNNA-based strategy outperforms the two strategies at all vaccination levels.
The number of iterations is $n_\mathrm{iter}=10^3$ for all cases except for network dismantling is $n_\mathrm{iter}=10^4$.
Because there is a small probability that weights get stuck in a local minimum, we took the best results out of eight trials.

\subsection{Transition of the optimal vaccination strategy in the SIRD model}

There is a crossover between the efficiency of the high-IFR and high-BC strategies in the SIRD model as illustrated in Fig.~\ref{fig:performance}(f).
Similar phenomena in the metapopulation model have been reported, and the first-order phase transition has been identified~\cite{Jhun2021a}; however, such research has not been extended to networks due to the lack of an appropriate method to study the optimal vaccination strategy in complex networks.
By considering the node features and tailoring the vaccination strategy to specific levels of vaccine supply, GNNA enables us to observe a new phenomenon in complex networks that could not be observed by the existing vaccination strategies.
Phi coefficient, which is identical to Pearson correlation coefficient for binary variables, of the optimal vaccination strategy identified by GNNA and BC/IFR-based strategy is illustrated in Fig.~\ref{fig:correlation}(a).
When only a small fraction of nodes can be vaccinated, the optimal strategy is similar to that of the high-IFR strategy.
However, when the vaccination rate approaches the point where herd immunity can be achieved, an abrupt transition occurs in the optimal vaccination strategy and involves vaccinating nodes with high BC.
The vaccination rate of the population divided into four age groups is illustrated in Fig.~\ref{fig:correlation}(b).
When the vaccination rate is low, the oldest age group 75--99, who has the highest fatality rate, is primarily vaccinated.
When the vaccination rate increases to approach herd immunity, the age group primarily targeted by the optimal vaccination strategy abruptly changes, and the population below age 50 is primarily vaccinated.
The senior age group is even less vaccinated than the rest of the population because they have a low contact rate (see Supplementary Figure S1).

\section{Conclusion}

We presented a vaccination framework based on GNNA, which can be implemented to minimize the damage, such as the total number of infections or epidemic mortality, caused by general epidemic processes.
The main advantage of GNNA is that it takes node features such as contagion, recovery, and fatality rate, and tailors the vaccination strategy to the level of vaccine supply available.
GNNA reduces the exponentially large dimension of the space of the vaccination to 7--10 and enables Gaussian random walk to effectively optimize vaccination strategies.
The efficacy and risk of vaccine side effects vary from individual to individual~\cite{Reingold1985,Kumar2008}.
GNNA can consider statistical estimation of such factors along with other risks (here, we only considered the age-dependency of the fatality rate) and morbidity.

We demonstrated that the optimal vaccination strategy is closely related to the total amount of vaccines available.
For instance, in the SIRD model, when vaccine supply is low, the optimal strategy primarily vaccinates nodes with high fatality rates, and when the vaccine supply is relatively high, it vaccinates nodes with high BC.
Such transition of the optimal vaccination strategy based on the vaccination rate can be identified by GNNA.
This transition is of theoretical interest also with real-world implications.
For instance, the hysteresis of the optimal vaccination strategy implies that mixing the fatality- and centrality-based strategies is ineffective in reducing the mortality rate~\cite{Jhun2021a}.
The proposed framework can be implemented in future research to find other new phenomena in the optimal vaccination strategies that couldn't be observed in the current centrality-based vaccination paradigm.
		
\begin{acknowledgments}
This research was supported by the NRF, Grant No.~NRF-2014R1A3A2069005.
\end{acknowledgments}

\end{document}


\title{Supplementary Material to \\Efficient vaccination strategy by graph neural vaccination}

\author{Bukyoung Jhun}
\email{jhunbk@snu.ac.kr}
\affiliation{CCSS, CTP and Department of Physics and Astronomy, Seoul National University, Seoul 08826, Korea}


\begin{table}
\begin{center}
\caption{\textbf{Number of nodes, number of edges, and epidemic parameters used for each network.}
}
\vskip 2mm
\setlength{\tabcolsep}{25pt}
\begin{tabular}{ccccc}
\hline
\hline
Network                       & $N$     & $L$       & $\beta$ & $\mu$ \\
\hline
Email                         & 1,103   & 5,451     & 0.1     & 0.5   \\
Airline                       & 3,304   & 19054     & 0.2     & 0.5   \\
Human contact multiplex            & 20,000 & 127,090 & 0.1     & 0.5   \\
Coauthorship: Complex Network & 32,016  & 100,434   & 0.2     & 0.5   \\
Coauthorship: DBLP            & 317,080 & 1,049,866 & 0.1     & 0.5   \\
\hline
\hline
\end{tabular}
\end{center}
\label{sm_tab:network_specifics}
\end{table}

\begin{figure}
	\includegraphics{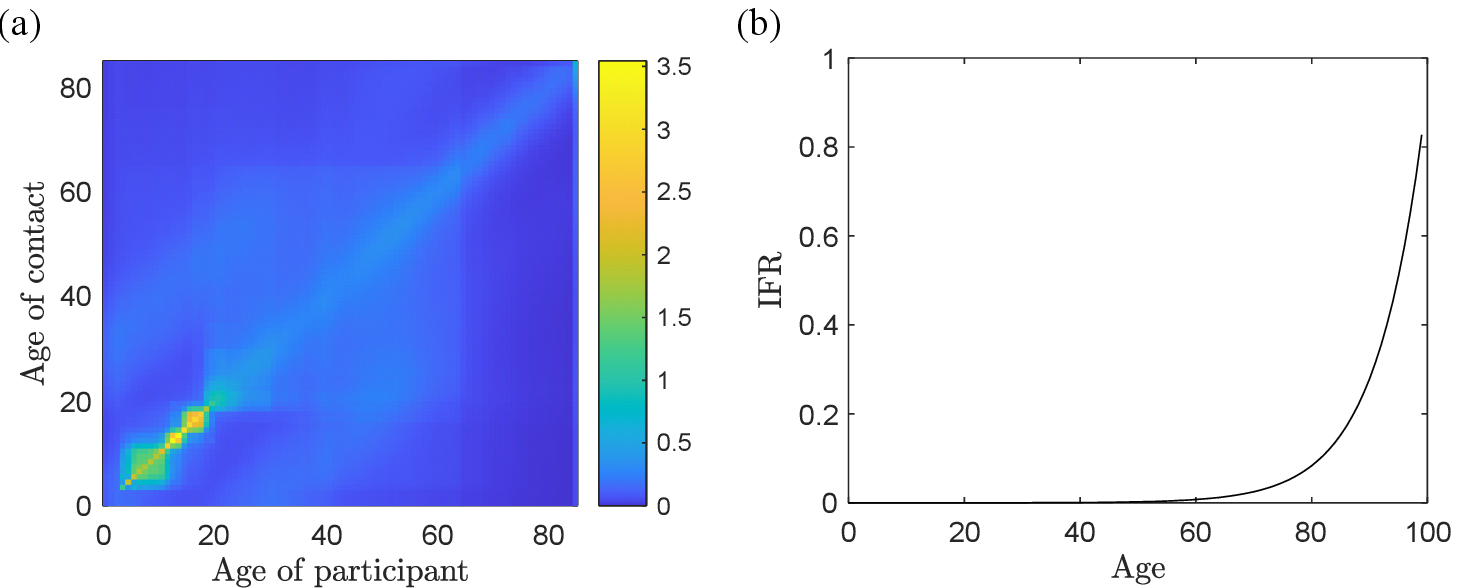}
	\caption{
	\textbf{Data used to construct the multiplex network for COVID-19.}
    (a) Contact matrix The contact rate between similar age groups is disproportionately higher.
    The senior population exhibits a low contact rate.
    (b) Age-dependent IFR was obtained from the meta-analysis of medical literature.
    }
\end{figure}

\begin{figure}
	\includegraphics[width=\textwidth]{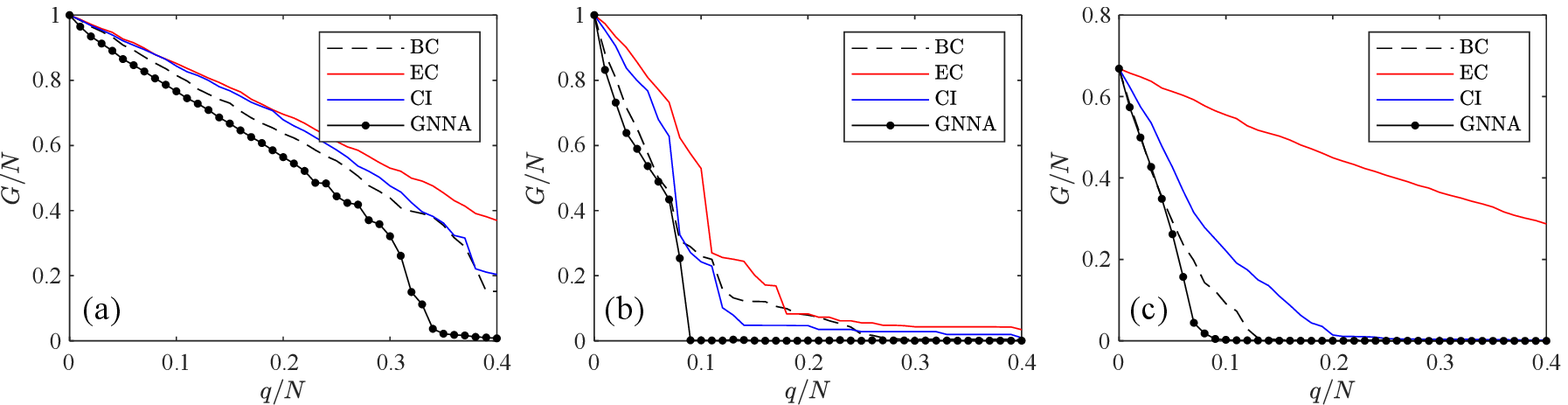}
	\caption{
	Size of the largest connected component of network dismantling problem.
	The strategies are tested on (a) email, (b) airline, and (c) coauthorship network of complex network research.
	The vaccination strategy obtained by GNNA outperformed all other centrality measures at all vaccination levels.
    }
    \label{sm_fig:dismantling}
\end{figure}

\begin{figure}
	\includegraphics[width=\textwidth]{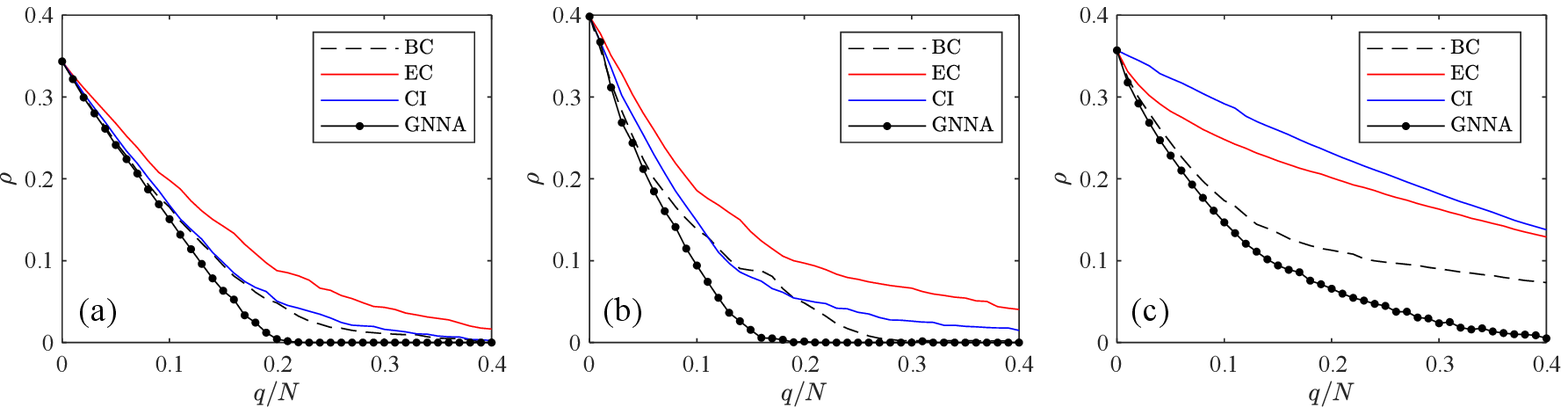}
	\caption{
	Density of infection of MMCA SIS model with homogeneous contagion and recovery rate.
	Te strategies are tested on (a) email, (b) airline, and (c) coauthorship network of complex network research.
	The vaccination strategy obtained by GNNA outperformed all other centrality measures at all vaccination levels.
    }
    \label{sm_fig:sis}
\end{figure}

\begin{figure}
	\includegraphics[width=\textwidth]{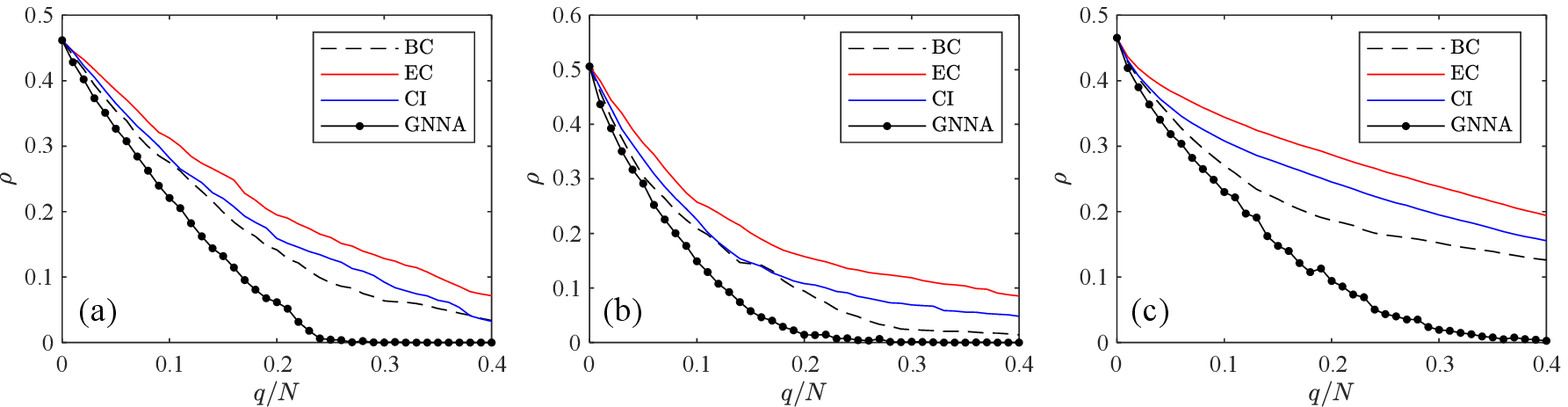}
	\caption{
	Density of infection of MMCA SIS model with heterogeneous contagion and recovery rate.
	The strategies are tested on (a) email, (b) airline, and (c) coauthorship network of complex network research.
	The vaccination strategy obtained by GNNA outperformed all other centrality measures at all vaccination levels.
    }
    \label{sm_fig:sis_hetero}
\end{figure}

\begin{figure}
	\includegraphics[width=\textwidth]{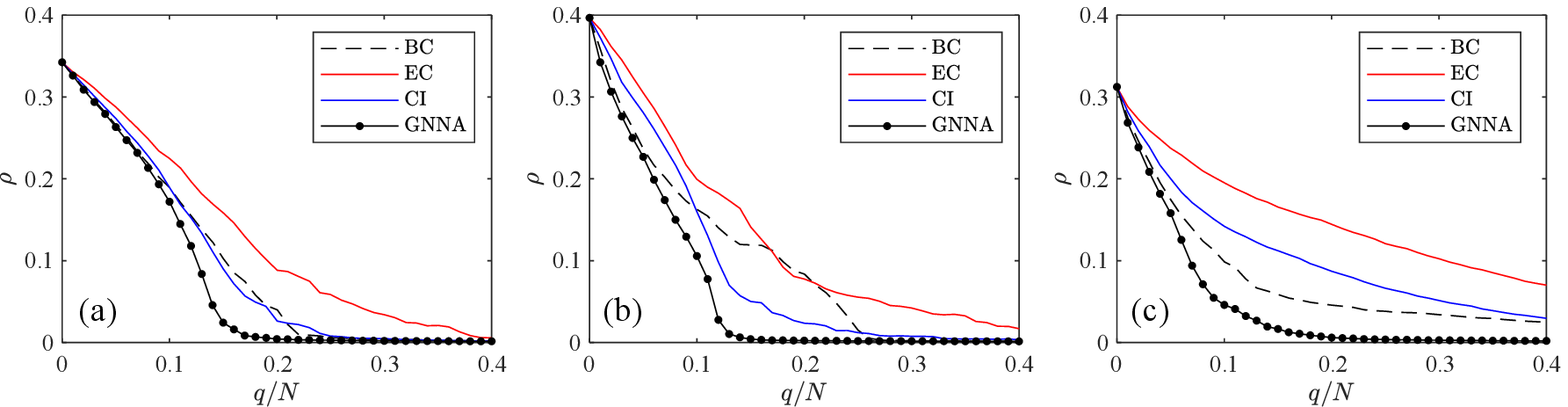}
	\caption{
    Density of infection of stochastic SIS model with homogeneous contagion and recovery rate.
	The vaccination strategy is obtained by the MMCA SIS model (depicted in Fig.~\ref{sm_fig:sis}).
	The strategies are tested on (a) email, (b) airline, and (c) coauthorship network of complex network research.
	The vaccination strategy obtained by GNNA outperformed all other centrality measures at all vaccination levels.
    }
    \label{sm_fig:sis_hybrid}
\end{figure}

\begin{figure}
	\includegraphics[width=\textwidth]{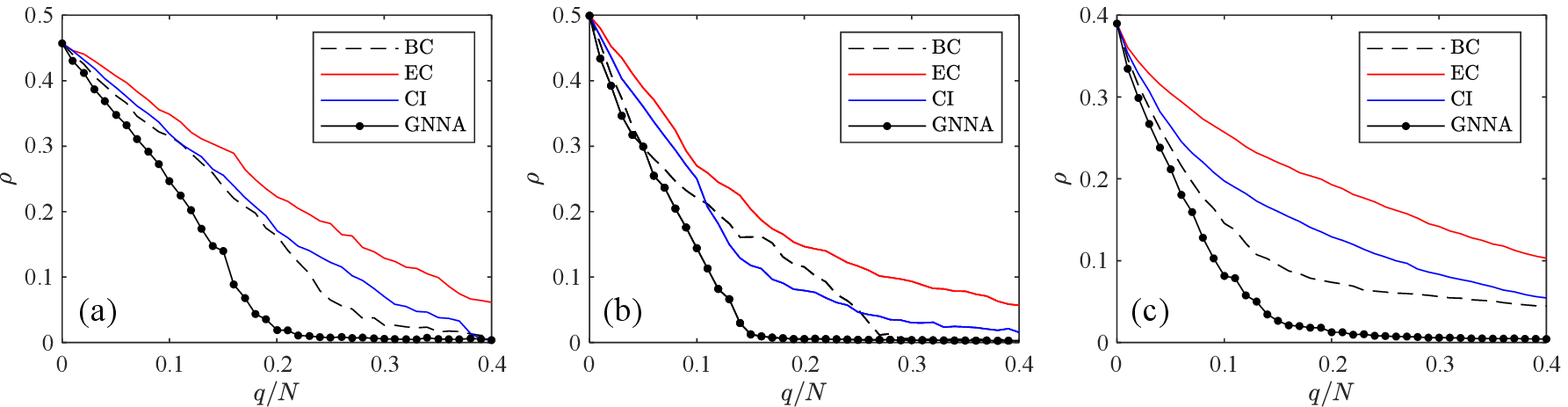}
	\caption{
	Density of infection of stochastic SIS model with heterogeneous contagion and recovery rate.
	The vaccination strategy is obtained by the MMCA SIS model (depicted in Fig.~\ref{sm_fig:sis_hetero}).
	The strategies are tested on (a) email, (b) airline, and (c) coauthorship network of complex network research.
	The vaccination strategy obtained by GNNA outperformed all other centrality measures at all vaccination levels.
    }
    \label{sm_fig:sis_hetero_hybrid}
\end{figure}